# A SHORT VIRTUAL REALITY MINDFULNESS MEDITATION TRAINING FOR REGAINING SUSTAINED ATTENTION

[1]ASATI Minkesh,  [2]MIYACHI Taizo

[1,2]Electrical and Electronics Engineering Department, Tokai University, Japan
Email: [1]minkeshasati@gmail.com, [2]miyachi@tokai-u.jp

**Abstract:** The ability to focus one's attention underlies success in many everyday tasks, but voluntary attention cannot be sustained for a long period of time. Several studies indicate that attention training using computer-based exercises can lead to improved attention in children and adults. A major goal of recent research is to create a short (10 minutes) and effective VR Mindfulness meditation particularly designed for regaining or improving sustained attention. In this study, we have created a custom virtually relaxing environment including an archery game with multiple targets. In the experiment, the attention span of 12 adults are tested before and after the virtual reality session by a non-action video game ([19]) score and Muse headband EEG-signals. After the 10-minute virtual reality session, participants' game scores increased (according to game experience): for the beginner by 275%, for intermediate by 107%, and for an expert by 17%. For Muse headband data, calm points increased by 250% irrespective of the participants gaming experience. After the experiment, all participants reported feeling recharged to continue their daily activities.

## 1. INTRODUCTION

Sustained attention is "the ability to direct and focus cognitive activity on specific stimuli over a long period of time." In order to complete any cognitively planned activity, any sequenced action, or any thought, one must use sustained attention. It is what makes it possible to concentrate on an activity for as long as it takes to finish, even if there are other distracting stimuli present. Examples of sustained attention include listening to a lecture, reading a book, playing a video game, or fixing a car. Problems occur when a distraction arises. A distraction can interrupt and consequently interfere in sustained attention.

Cohen R.A. (2011) [1] describes Sustained attention as one of the primary elements or component processes of attention. It enables the maintenance of vigilance, selective and focused attention, response persistence, and continuous effort despite changing conditions. DeGangi and Porges (1990) [2] indicate there are three stages to sustained attention which include: attention getting, attention holding, and attention releasing. The level of sustained attention varies from person to person. However, a key aspect of sustained attention is the ability to refocus on the task after a distraction arises.

Mindfulness meditation training has been linked to a broad range of cognitive, affective, and health outcomes. Some of the most robust findings in the cognitive domain pertain to how mindfulness meditation training can foster on-task, sustained attention and reduce mind-wandering (see e.g. [3] [4] [5] [6] [7] [8]). Hayley A. Rahl,2016 [9] tested two competing accounts for how mindfulness training reduces mind-wandering.

Virtual reality technologies have been successfully used in many therapies, especially those that rely on mental imagery: to elicit and modulate psychophysiological symptoms of anxiety and fear reactions, in both patients with anxiety disorders and healthy individuals [10]. Patients with phobias who traditionally have attempted to desensitize themselves within their imagination can face their fears in a controllable virtual environment [11]. Similarly, patients with eating disorders who suffer from distorted body image can change their self-image through a virtual body [12].

We propose "10-minute Virtual Reality Mindfulness Meditation Training" particularly for regaining sustained attention. Anyone can take even in the middle of busy everyday life for recharging and regaining sustained attention at their best.

## 2. EXISTING METHODOLOGIES AND THEIR PROBLEMS

### 2.1 Most effective existing methodology/system of Sustained Attention Training

Traditionally, meditation has been considered as something that needs to be exercised almost daily and for long periods. However, recently there has been a growing interest in short-duration meditation or mindfulness programs, which could provide results quickly. There are now several studies showing that brief mindfulness meditation training reduces mind-wandering and improve sustained attention (see e.g. [4] [5]).

Neuroadaptive virtual reality meditation system (that combines virtual reality with neurofeedback) provides a very effective Meditation and Mindfulness Training. Shaw et al. [13] introduced the Meditation Chamber, an immersive virtual environment for meditation training. The system used skin conductance as the biofeedback mechanism in three guided meditation and relaxation exercises. RelaWorld [14] measures participants' brain activity in real time via EEG and calculates estimates for the level of concentration and relaxation. These values are then mapped into the virtual reality. Similarly, PsychicVR [15] non-invasively monitors and records the electrical activity of the brain and incorporates this data in the VR experience using an Oculus Rift and the MUSE headband. When the participant is focused, they are able to make changes in the 3D environment and control their powers (focusing ability).

A study by Zeidan et al. [16] hints that a brief mindfulness meditation intervention of only three sessions leads to reduced heart rate and increased heart rate variability (which is related to well-being and positive affect; [17]) immediately after or during meditation tasks, whereas Steffen and Larson [18] present evidence of reduced cardiovascular reactivity to a stressor during a single mindfulness meditation session.

Traditional approach (see e.g. [3] [4] [5] [6] [7] [8]) are effective but not enough attractive to motivate the participant to take the session in the busy daily schedule. Neuroadaptive virtual reality meditation training is enjoyable and effective but time-consuming (EEG or Skin conductance sensors require at least 10-minutes to record quality signals before starting the training). Also, these are complex procedures such as attaching EEG sensors or skin conductors to the body and those are not usually comfortable during the mediation. To minimize the downsides of these systems, in this research, we are contributing a short and simple but effective VR mindfulness meditation training particularly for improving or regaining sustained attention.

## 2.2 Problems of existing methodology/system of Sustained Attention Training

P1: Hand-eye coordination training is not available
Hand-eye coordination is a complex cognitive ability, as it calls for us to unite our visual and motor skills, allowing for the hand to be guided by the visual stimulation our eyes receive. Most activities that we do in our day-to-day life use some degree of eye-hand coordination, and it is a crucial aspect of sustained attention training.

P2: Reaction time training is not available
Reaction time refers to the amount of time that takes places between when we perceive something to when we respond to it. It is the ability to detect, process, and respond to a stimulus. Reaction time necessarily includes a motor component, unlike processing speed. Therefore, having good reaction time is associated with having good reflexes. Its mean distractions can be avoided without dividing the attention.

P3: Traditional video game is not enjoyable
Some research is effective, but they are based on traditional computer/mobile video games (action or non-action) training. These video games are not interesting enough in this modern VR world to motivate the participant to play it.

We explained above three problems (P1, P2, P3) where P1 and P2 belongs to existing VR mindfulness mediation systems ([13] [14] [15]) and P3 belongs to traditional approaches ([3] [4] [5] [6] [7] [8]).

## 3. Virtual Reality Mindfulness Meditation Training

### 3.1 Methodology for problems
Solution 1 for P1:
We introduced a hand-eye coordination training in the VR archery game by placing five targets in a very systematic position (four of them are placed on the corners and one is in the center of a virtual cube). Additionally, the archer must shoot in a way similar to shooting in real life archery; otherwise the archery shot will not be complete. In these situations, hand-eye coordination is very important.

Solution 2 for P2:
We also included reaction time training in the VR archery game. If the arrow hits the target, then it will flash out (and disappear) and reappear after five seconds. There are total 5 targets so that participant need 5 seconds to finish all targets. If the participant takes one second to finish one target then after 5 seconds, the participant will see only one target and it will be continuing until participant continue this cycle (one target in one second). Participants are advised to finish all targets in a way that you should be able to see only one target at a time. To achieve this condition, the archer must shoot each target in one or less than one second. In these situations, the reaction time is going to play an important role.

Solution 3 for P3:
An immersive and calm virtual environment was created where the participant can walk and will feel refreshed by observing the mountains, trees, fireplace, paintings, and aquarium.

### 3.2 Strong Points of 10-minute Virtual Reality Mindfulness Meditation Training
1. Hand-eye Coordination Training with archery game in a very calm and enjoyable virtual environment.

2. Reaction time training by creating a competitive situation in the archery game.
3. Peaceful music, beautiful scenery, and realistic feeling of air flow through the mountains and jellyfish movement in an aquarium to give an immersive and refreshing virtual experience to the participant.

These three points together make an effective VR Sustained Attention Training through Mindfulness Meditation session.

### 3.3 Our Method vs Other Archery games

Ordinary archery games are designed just for fun or general brain training. Our archery game is particularly designed for sustained attention training, and VR immersion provides an enjoyable and calm environment. The main advantages are as follows:

(1) Easy shooting:
The participant does not have to do any extra effort to get/align the arrow with the bow because the arrow will appear and be aligned automatically when the archer pulls the string of the bow.

(2) Easy adjustment with short distance motion to next target:
Archery game has six-degree of freedom. Therefore, the participant can easily change their short distance position and orientation to shoot the next target.

(3) Rhythmical shooting Motivation:
In the archery game, there are total 5 targets. If the arrow hits the target, then it will flash out (and disappear) and reappear after five seconds. This will create a cycle and motivate the archer to keep focus and continue the cycle (one target in one or less than one second) because if a single target is missed, then the cycle will be broken.

### 3.4 System Feature and Structure

The underlying motivation and design principle of this system was to create a short training by combining the proven methods of Mindfulness Meditation and virtual reality into one package that would allow novice or even experienced meditators to regain the sustained attention even in the middle of busy everyday life.

Virtual Reality Environment:
A Virtual Reality Environment was created with a room made of three walls. Instead of the 4th wall, there was a balcony, and from the balcony, the participant a wonderful view of mountains and trees. To decorate the room, there was a fireplace, some paintings, and chandeliers. We created a large aquarium in the balcony. By standing beside the aquarium, participant can shoot light-bubbles (archery target) with an archery bow. In the aquarium, there are some jellyfish, making the real sound while moving to provide a calm environment to the participant. Also, the participant can feel the sound of air and tree leaves while standing or playing the archery game in the balcony. This setup is created to provide a very calm environment and fresh feeling for the participant.

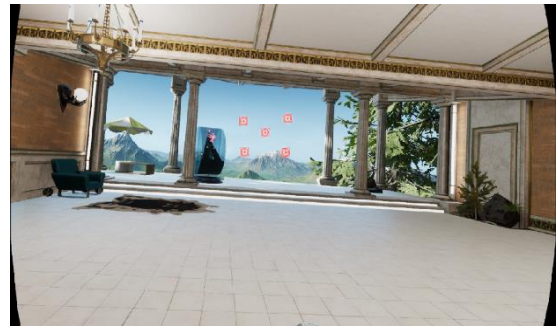

*Figure 1: Immersive VR environment (Scene)*

Hand-eye Coordination Training:
In the archery game, the participant is standing beside the aquarium, oriented towards the mountain view. There are five light bubbles (archery target). Four of them are placed on the corners, and one is in the center of a cube. The archer does not have to do any extra effort to get/align the arrow with the bow because the arrow will appear and be aligned automatically when the archer pulls the string of the bow. However, both of the archer's hands should be aligned in the right direction (such as the real archery hand positions); otherwise the shot will not be completed. This type of situation is created for Hand-eye coordination Training.

Reaction Time Training:
In the archery game, if the arrow hits the target, then it will flash out (and disappear) and reappear after five seconds. Participants must finish all target in a way where you are able to see only one target at a time. To achieve this condition, the archer must shoot a target in one second or less. This will create a cycle and motivate the archer to keep focus and continue the cycle because if a single target is missed, then the cycle will be broken, and more than one target will have appeared. These situations are created for Reaction Time Training.

This training helps the participant to increase the attention span, and Virtual Reality Environment provides an enjoyable and calm environment. All these three parts together make an effective sustained attention training.

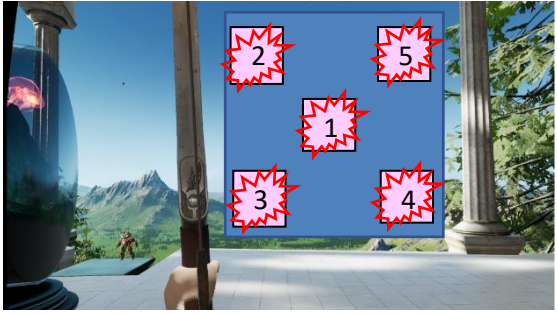

*Figure 2: VR Archery game (1 bow and 5 target)*

### 3.4 Evaluation

To evaluate the system, we were looking for a game which would be able to check response time and sustained attention with a very good participant experience. Finally, we selected a non-action video game [19] after playing more than 30 similar types of games. All you must do in this game is guess the right color and tap on the right answer of the three given options. Hence the questions are a set of words spelling a color, although the color in which they are written is different. You must select the right color filling that word. What makes this game interesting but confusing at the same time is that when you play the game, there is a strict time limit for each answer. Therefore, it becomes very difficult, and any break in your attention forces you to start over. Your score increases with every right answer you give, and you start over every time you answer incorrectly.

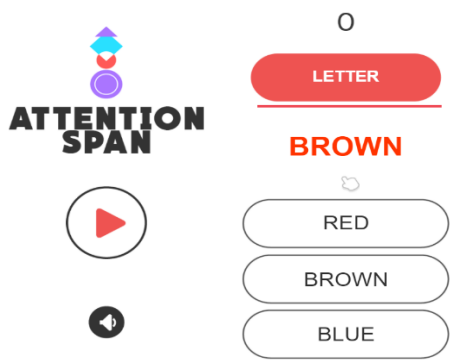

*Figure 3: Home and Play screen of Game [19]*

We let participants practice the video game [19] before evaluation to become familiar with it. In the evaluation process, participants were instructed to play this game before (three times) and after (three times) the VR Training and recorded the game-score. We set up to play the game three times before and after the training to get a precise score (average) and reducing any accidental errors. Also, participants wore Muse headband (5 EEG sensors) while playing the game, and we captured the mind state of the participant by recording the eye blinking and EEG signals. Data was recorded in the interval of every 100 ms (10 times in a second).

## 4. EXPERIMENTS

### 4.1 Participants
Subject:
12 random participants (eight male and four female) were selected between 20 to 24 years old.

During selection, we also considered gaming experience of the participants (four beginning, four intermediate, and four expert gamers).

Two participants wore glasses (-1<eyesight<+1.25) on daily basis. Also, participants were invited to our lab in the middle of the day when their morning freshness has faded, and their sustained attention is not at their best.

Major arm:
Two participants were left handed. The rest of the 10 participants were right handed.

Game name:
VR Archery for sustained attention.

### 4.2 Procedure:
When arriving in the lab, the participant sat in a comfortable chair. We checked the heart rate with a wristband (Fitbit charge 3) and waited until the heart rate is not steady to get a real insight of heart rate variation during the experiment (training and evaluation).
(1) VR participants experienced a demo game (Oculus first contact) before VR training.
(2) Before and after the VR Training, the participant is instructed to play a computer-based non-action video game [19] to evaluate the attention span.
(3) We let the participant practice the video game [19] before evaluation for getting familiar with it. For evaluating the calmness of the participant during the evaluation game, a Muse-headband were used to record brain signals and eye-blink data.
(4) The participant experienced the Mindfulness session (sustained attention training) for 10 minutes.
(5) After the experiments, we thanked the participant and provided a short briefing about the purpose of the experiment.

The whole experiment session took between 25 to 40 minutes.

### 4.3 Hardware
This Mindfulness Meditation Session is a 3D-virtual reality environment designed for Sustained Attention Training. It utilizes the Oculus

Rift DK2 head-mounted display and its touch controllers. A Fitbit charge-3 wristband used to check the heart rate variability of the participant during VR Training and Evaluation. A Muse headband was used to capture eye-blink data and EEG signals during evaluation.

## 5. RESULTS AND DISCUSSIONS

**5.1 Results:**

Every participant played the game [19] for evaluation before (**3-times**) and after (**3-times**) the training and recorded the scores. Also, by using the Muse headband, three types of data were recorded, a) Calm points, b) Recoveries, and c) Birds count. Also, by using a wristband (ECG sensors), heart rate variability (HRV) was recorded during the entire experiment (training as well as evaluation).

In this section, we define all the terms in which we have presented the recorded muse data. Participants' state of mind can be divided into three states (neutral, calm, and active). Calm points are calculated with this formula – [(neutral (seconds) * 1 + calm (seconds) * 3)/8]. Recoveries mean total counts of returning from the active state to the neutral state. Also, birds count stands for being in a calm state for a long time.

Evaluation game [19] is an endless game. So, for getting the highest score possible, we let 3 participants (1 from each category) to play the game until they are satisfied with their score and picked the highest one among all game's scores. They played the game around 45-50 times to achieve these scores.

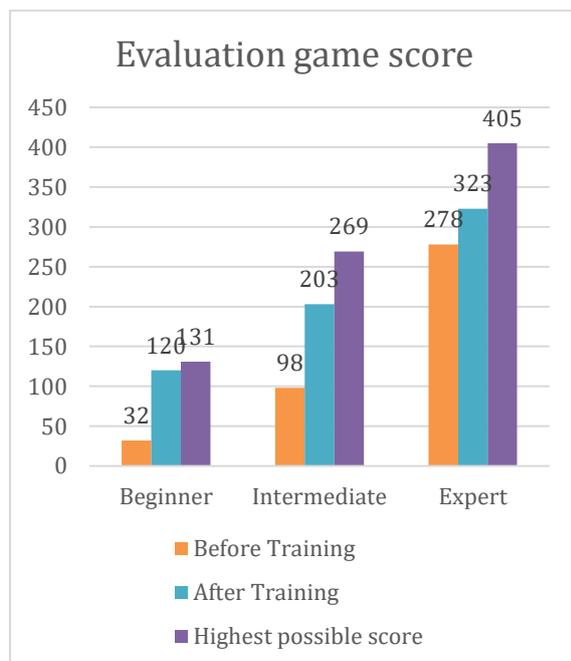

*Figure 4: Average game score for categories. Highest possible score is the absolute score.*

Game [19] score increased on average for the 12 participants: for the beginner from 32 to 120, for intermediate from 98 to 203, and for the expert from 278 to 323, after the training.

Also, by observing three types of data recorded by the Muse headband, calm points increased from 6 to 21, Birds count increased from 3 to 10, and Recoveries decreased from 2 to 0 – indicating that participants always remain either in neutral or in a calm state (did not enter the active state) after the training. Headband data was irrespective of the participants' gaming experience.

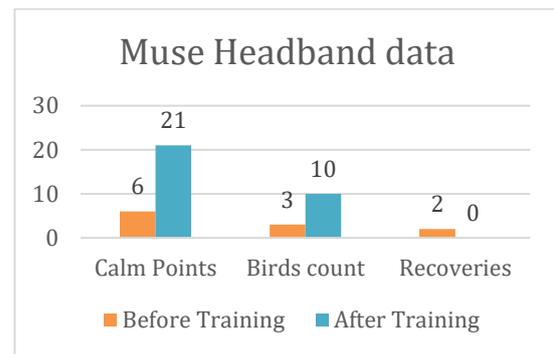

*Figure 5: Average data of all participants*

We found that participants' game scores increased for beginner by 275%, for the intermediate by 107%, and for expert by 17%. For Muse headband data, calm points increased by 250%, Birds count increased by 233%, and Recoveries decreased from two to zero. Participants were able to regain sustained attention at their best and recharge for the rest of the day. Heart rate variability (HRV) also slightly increased after the experiments.

**5.2 Discussions**

Evaluation game [19] required sustained attention and even mind wandering for a second can force participant to restart the game. After the training, participants recharged and got their sustained attention at their best that's why they were able to perform better. As well as Muse-headband data also shows that they were fully calm, and their mind remained always in a calm or neutral state during the evaluation game.

In this section, we will describe heart rate changes during the VR training. In the beginning, when participants were observing the immersive virtual environment, heart rate increased but slowly went down to normal as participant got familiar with the Virtual environment. Moreover, during the VR archery game, heart rate was constantly increasing as the participant came close to achieving the game's objectives and then changes were persistent during the entire time of the game. We can relate these changes with the competency of the game (for response time training), and these changes will always be there whenever participant plays the game. Because of this, we are considering these changes as an improvement in HRV (Heart Rate Variability). Furthermore, during the evaluation after the training, heart rate was near to the resting heart rate of the participant. That indicates that participants were so relaxed and focused after VR training.

Poor vision can affect the VR experience of the participant. In our experiments, three participants use glasses on a daily basis. The first one had the eyesight of +1.25 for the right eye and +1.00 for the left eye. The second one had the eyesight of +0.75 for the right eye and +1.00 for the left eye. The third one had -6.00 for the left eye and -5.00 for the right eye. We asked the participants to close their eyes one at a time. We discovered that eyesight of in between ±1 don't affect the experience of the participant in VR. They shared the experience as a normal participant. However, more the +1 and less than -1 affected the experience at a significant level. That's why we didn't consider the 3rd participant evaluation scores in our average.

In general, high game scorers cannot improve their scores after any type of short training. Also, in our experiments expert game participant scores increased by only 17% even though they felt refreshed and their biological data improved (improved sustained attention and reduced mind-wandering) compared to a normal participant. That's where we were able to achieve our objective of improving sustained attention and recharging the participant for the rest of the activity of the day after only a short 10-minute training.

In general, left-handed people do better in games. In our experiments, left-handed (two participants) participants because familiar with the Oculus touch easily, and their target hit rate (archery game) was also slightly better than right-handed participants.

When we asked participants about the difference between this VR archery game and other similar types of games (VR or Non-VR), most of them replied that the game was enjoyable and even they can play it daily because of combination of competency, VR Immersive environment, hand-eye coordination training, and ease of use.

## 6. CONCLUSION

In this research, we are contributing a short (10 minute) and effective VR Mindfulness meditation particularly designed for regaining or improving sustained attention. Anyone can use it in the middle of a busy day to recharge and regain sustained attention at their best. We evaluated our training by a game [19] score, Muse headband data, and ECG sensor data. Participants mind wandering (no one entered the active state) was significantly reduced, and evaluation-game [19] score was increased after the training.

## 7. ACKNOWLEDGMENTS

We are grateful for the support of the Japan International Agency (JICA).